\documentstyle[multicol,prl,aps,epsf]{revtex}
\begin{document}

\title{On the Scaling Behavior of  the Abelian Sandpile Model}

\author{Barbara Drossel}
\address{Department of Physics, 
  University of Manchester, Manchester M13 9PL, England}
 
\maketitle

\begin{abstract} 
The abelian sandpile model in two dimensions does not show the type of
critical behavior familar from equilibrium systems. Rather, the
properties of the stationary state follow from the condition that an
avalanche started at a distance $r$ from the system boundary has a
probability proportional to $1/\sqrt{r}$ to reach the boundary. As a
consequence, the scaling behavior of the model can be obtained from
evaluating dissipative avalanches alone, allowing not only to
determine the values of all exponents, but showing also the breakdown
of finite-size scaling.

\end{abstract}

\begin{multicols}{2} 

Since its introduction in 1987, the sandpile model has been considered
as the prototype of a self-organized critical (SOC) system
\cite{bak87}. Computer simulations suggest that irrespective of the
initial conditions and of details of the model rules, the system self
organizes into a ``critical'' state with a power-law size distribution
of avalanches. The concept of SOC is thought to explain the frequent
occurrence of power laws in nature.

Considerable effort has been made to determine the scaling behavior of
the two-dimensional abelian sandpile model. The advantage of this
model is that several of its properties can be calculated analytically
\cite{dha90,iva94}. However, the exponent $\tau_r$ characterizing the
distribution of avalanche radii, and related exponents for the
distribution of the avalanche duration $t$ and the number of topplings
$s$ during an avalanche have resisted any attempt of an analytical
calculation. The numerical determination of these exponents is
hampered by the fact that the double logarithmic plots show slopes
that increase with increasing system size, until finite-size effects
set in, indicating that the asymptotic scaling behavior does not yet
occur for the system sizes accessible to computer simulations. Thus,
the predictions for the value of $\tau_s$ vary between 1.22
\cite{manna90} and 1.27 \cite{che99} or even 1.29 \cite{lub97}. The
latter two results where obtained under the assumption that the system
displays finite-size scaling (FSS).

Recently, evidence was found that the abelian sandpile model does not
display FSS \cite{teb99}. This result was obtained from an
investigation of multifractal spectra. Indeed, there is no a priori
reason why the abelian sandpile model should display FSS. The concept
of FSS has its origin in equilibrium critical phenomena where a small
finite system cannot be distinguished from a small part of a large
system. However, this is not the case for the abelian sandpile model
in two dimensions. In a small system, the sites that participate in an
avalanche may topple a few times during the duration of the
avalanche. In a larger system, the number of topplings per site during
a large avalanche is larger. Thus, locally collected data (i.e. the
number of topplings of a given site) contains information about the
size of the system. Also, finite-size scaling is based on the
assumption that boundaries play no special role in the system.
However, the boundaries of the abelian sandpile model play an
essential role as they are the only place where sand can leave the
system.

It is the purpose of this letter to elucidate the scaling behavior of
the abelian sandpile model. Since boundaries play a special role in
the system, it is essential to consider avalanches that reach the
boundaries of the system separately from those that don't. It turns
out that the dissipative avalanches display beautiful power
laws. These power laws imply that each avalanche started at a distance
$r$ from the system boundary has a probability proportional to
$1/\sqrt{r}$ to reach the boundary and, if it does so, dissipates on
an average $\sqrt{r}$ sand grains. From the scaling behavior of
dissipative avalanches, we obtain the values of the critical exponents
characterizing the system, and we see compelling evidence for the
violation of FSS.

The two-dimensional abelian sandpile model is defined on a square
lattice with $L^2$ sites. At each site $i$ an integer variable $z_i$
represents the number of grains. Grains are added individually to
randomly chosen sites of the system. When the number of grains at a
site $i$ exceeds $z_c=3$, site $i$ is unstable and topples, its height
being reduced to $z_i-4$, and the heights $\{z_j\}$ of all four nearest neighbors
being increased by one. If $i$ is a boundary site with $l<4$
neighbors, $4-l$ grains leave the system. If a neighbor $j$ becomes
unstable due to the addition of a grain, it also topples, and the
avalanche stops when a new stable configuration is reached. During an
avalanche, no new grains are added to the system.  The size $s$ of an
avalanche is defined to be the total number of topplings. The radius
of an avalanche is in this paper taken to be the maximum distance of
toppled sites from the starting site of the avalanche. 
It has proven useful to decompose an avalanche
into ``waves of toppling'' \cite{pri96}. The $n$th wave of toppling begins when the
starting point of an avalanche topples for the $n$th time, and all
those sites belong to it that topple immediately after a nearest
neighbor that belongs to the same wave has toppled.

\begin{figure}
\centerline{{\epsfysize=0.3\columnwidth{\epsfbox{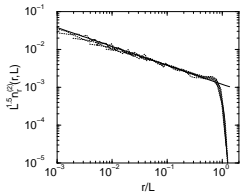}}}}
\narrowtext{\caption{The radius distribution of avalanches that touch the boundaries of the system, for $L=$128, 256, 512, 1024, 2048. The thick solid line is a power law with an exponent -1/2.\label{fig1}
}}
\end{figure}

Let us first discuss the radius distribution of avalanches,
$n_r(r,L)$.  Since its asymptotic (i.e., $L\to \infty$) behavior is
difficult to extract from simulation data, recent results for the
exponent $\tau_r$ vary between $7/5$ \cite{teb99} and $5/3$
\cite{lub97}.  In \cite{men98}, a version of the abelian sandpile
model was studied where sand grains were only dropped at the center of
the system. It was found that the fraction $1/\sqrt{L}$ of all
avalanches reach the boundaries. This finding would imply that
$\tau_r=3/2$, just as predicted in \cite{pri96}. This is indeed the correct exponent, as a separate
evaluation of dissipative and nondissipative avalanches shows. 
We denote by $n_r^{(1)}(r,L)$ the radius distribution of nondissipative avalanches, i.e., those avalanches that do not reach the boundaries. The distribution of dissipative avalanches is denoted $n_r^{(2)}(r,L)$. Normalization of the radius distribution requires
$$\int_1^\infty [n_r^{(1)}(r,L)+n_r^{(2)}(r,L)]dr = 1.$$ 
Fig.~\ref{fig1} shows the result of computer simulations for $n_r^{(2)}(r,L)$.
The distribution $n_r^{(2)}(r,L)$ has the form
\begin{equation}
n_r^{(2)}(r,L) \sim L^{-3/2} g(r/L) \label{n2}\,.
\end{equation}
The prefactor is such that the total weight of $n_r^{(2)}(r,L)$ is $\int_1^L
n_r^{(2)}(r,L)dr \sim 1/\sqrt{L}$, in agreement with the result in \cite{men98} that the
fraction $1/\sqrt{L}$ of all avalanches reach the boundaries.  Since
grains are dropped randomly into the system, the probability that an
avalanche is triggered at a fixed distance $r \ll L$ from the boundary
is proportional to $1/L$, leading to $n_r^{(2)}(r,L) = (1/L) f(r)$ for $r \ll
L$. Together with Eq.~(\ref{n2}), this gives $g(r/L) \propto
(r/L)^{-1/2}$ for small $r/L$. From Fig.~\ref{fig1}, it can be seen
that this $1/\sqrt{r}$-behavior extends almost over the entire range
of $r$ values. An avalanche triggered at a distance $r$ from the
boundary consequently has a probability proportional to $1/\sqrt{r}$
to reach the boundary, and if it reaches the boundary it dissipates on an average of the order $\sqrt{r}$ sand grains. 

\begin{figure}
\centerline{{\epsfysize=0.6\columnwidth{\epsfbox{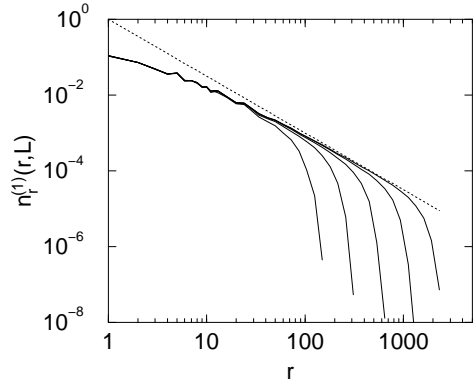}}}}
\narrowtext{\caption{The radius distribution of avalanches that do not touch the boundaries of the system, for $L=$128, 256, 512, 1024, 2048 from left to right. The dotted line is a power law with the exponent $-3/2$.\label{fig2}
}}
\end{figure}

Fig.~\ref{fig2} shows the radius distribution of avalanches that do
not reach the boundaries.  For the system sizes studied in the
simulations, no power law is visible. The slope seems to become
steeper with increasing $r$. (The last part of each curve, which
begins at the point where it splits from the curves for larger system
sizes, is due to finite-size effects and it must be ignored in the
subsequent discussion which refers to the thermodynamic limit $L\to
\infty$.) However, the curves $n_r^{(1)}(r,L)$ must approach a power
law $n_r^{(1)}(r,\infty)\sim r^{-\tau_r}$ for sufficiently large
$r$. The reason is that the last wave of topplings of an avalanche has
been proved to be distributed according to a power law $ r^{-7/4}$
\cite{dha94}, and $n_r^{(1)}(r,\infty)$ cannot be steeper than
this. From the above results for $n_r^{(2)}(r,L)$, we can even deduce
the value of $\tau_r$: The probability that an avalanche triggered at
distance $r$ from the boundary reaches the boundary is proportional to
$Ln_r^{(2)}(r,L)$, and it is identical to the probability that an
avalanche triggered in the interior of the system reaches at least a
radius $r$, which is given by $\int_r^\infty
n_r^{(1)}(r',\infty)dr'$. The reason is that the landscape of height
values in the abelian sandpile model does not show long-range
correlations. Rather, correlations decay as fast as $r^{-4}$
\cite{dha90}, implying that avalanches spread in the same way
everywhere in the system, as long as they do not encounter the
boundaries.  Consequently, $n_r^{(1)}(r,\infty)$ must fall off as
$r^{-3/2}$ for sufficiently large $r$, fixing the value of
$\tau_r=3/2$. Fig.~\ref{fig2} shows that this asymptotic value is
barely reached for the system sizes accessible to computer
simulations. This explains why evaluations based on the complete
radius distribution $n_r^{(1)}(r,L)+n_r^{(2)}(r,L)$ do not reveal the
correct exponent.

The distribution of the number of topplings $n_s(s,L)=n_s^{(1)}(s,L)+n_s^{(2)}(s,L)$ in
avalanches can be analyzed in a similar way. The superscript $(1)$ refers again
to nondissipative, the superscript $(2)$ to dissipative avalanches. If the
number of topplings $s$ in an avalanche of radius $r$ was always of the order $r^D$, simple
scaling would hold, resulting in a FSS form $n_s(s,L) \sim
s^{-\tau_s} {\cal C}(s/L^D)$ and a scaling relation
$D=(\tau_r-1)/(\tau_s-1)$. The breakdown of simple scaling was pointed out for the first time in \cite{men98,teb99}. It can best be
visualized by looking at the distribution of the number of topplings
$n_s^{\rm (spanning)}(s,L)$ of those avalanches that span the entire
system, i.e.~that touch all four edges of the boundary. These
avalanches have a radius of the order $L$. 
\begin{figure}
\centerline{{\epsfysize=0.6\columnwidth{\epsfbox{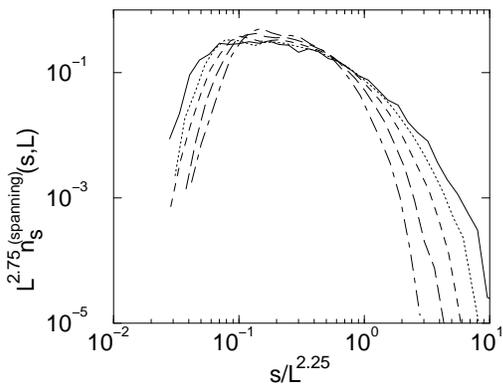}}}}
\narrowtext{\caption{The distribution $n_s^{\rm (spanning)}(s,L)$ of the number of topplings in system-spanning avalanches for $L=$ 2048,1024,512,256,128 (from widest to narrowest curve). 
\label{fig3}
}}
\end{figure}
Evaluation of the curves represented in Fig.~\ref{fig3} shows that
the maximum scales as $L^D$ with $D=9/4$, the lower cutoff as $L^2$,
and the upper cutoff as $L^{11/4}$. These exponents are obtained by
scaling $s$ in such a way that the maxima, the left tails, or the
right tails collapse. The mean scales roughly as $L^{5/2}$. 
Clearly, a complete scaling collapse would only
be possible if the four quantities were characterized by the same
exponent. 
The scaling behavior of the mean implies 
$$ \bar s =\int_1^\infty s n_s(s,L)ds \sim L^2,$$
in agreement with the analytical result given in \cite{dha90}.

\begin{figure}
\centerline{{\epsfysize=0.6\columnwidth{\epsfbox{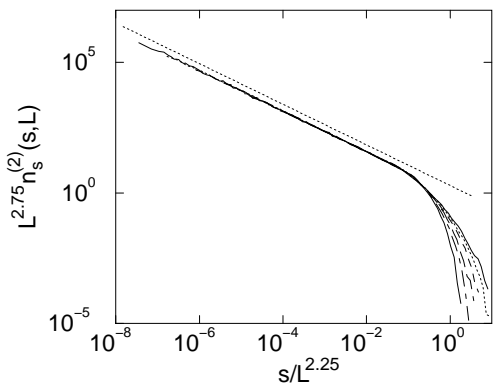}}}}
\narrowtext{\caption{The distribution $n_s^{(2)}(s,L)$ of the number of topplings in dissipative avalanches for $L=$ 2048,1024,512,256,128,64 (from widest to narrowest curve). The dotted line is a power law with the exponent -7/9.
\label{fig4}
}}
\end{figure}

Fig.~\ref{fig4} shows the distributions $n_s^{(2)}(s,L)$ of the
number of topplings in dissipative avalanches.
The curves have the form $n_s^{(2)}(s) \sim L^{-1}s^{-7/9}$ for
sufficiently small $s$, and then have a cutoff that does not display
FSS but is similar to the one for spanning avalanches.  The total
weight of dissipative avalanches, $ 1/\sqrt{L}$, is proportional to
$\int_1^{L^{9/4}}n_s^{(2)}(s,L)ds$.  The exponent $\tau_s$ can be
derived from the condition
$$s^{1-1/D}Ln_s^{(2)}(s,L)\sim \int_{s}^\infty n_s^{(1)}(s',\infty)ds'\, ,$$
giving $\tau_s=11/9$, which agrees very well with the value obtained
by Manna \cite{manna90}. The asymptotic slope $-\tau_s$ is
barely visible in $n_s^{(1)}(s,L)$ for the system sizes accessible to computer
simulations. For the value of the exponent $\tau_s$, the multiscaling
features are irrelevant, since only the exponent $D=9/4$ enters the
evaluation of $\tau_s$. However, the multiscaling behavior is relevant for the result $ \bar s = L^2$, which stands in no relation to the value of $\tau_s$. 

\begin{figure}
\centerline{{\epsfysize=0.6\columnwidth{\epsfbox{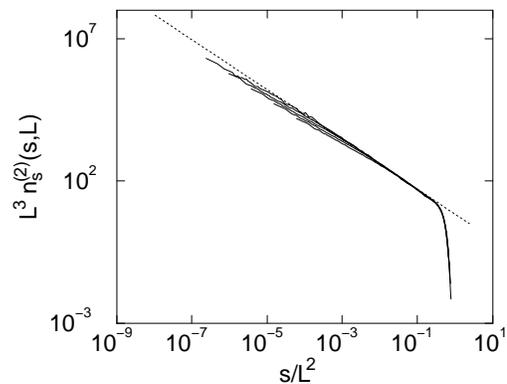}}}}
\narrowtext{\caption{The distribution $n_s^{(2)}(s,L)$ of the number of topplings in dissipative first waves for $L=$ 2048,1024,512,256,128. The dotted line is a power law with the exponent -7/8.
\label{fig5}
}}
\end{figure}

The failure of FSS in the system is due to a broad distribution of the number of waves of toppling in avalanches. Single waves of toppling display FSS, as shown in Fig.~\ref{fig5} for the first wave of toppling. 
The scaling with $L^2$ follows directly from the fact that waves are
compact and that each site topples once in a wave. Repeating the
evaluation performed already twice, we find an exponent $\tau_s=11/8$
for the first wave, which is identical to the analytically derived
exponent for the last wave \cite{dha94}, and different from the
exponent $5/4$ for boundary avalanches in a sector of $360^0$
\cite{iva94}. From the condition that waves are compact, it follows
immediately that $\tau_r=7/4$ for the first wave, which means that a
fraction proportional to $1/L^{3/4}$ of all first waves reach the
boundary. The result $\tau_r=7/4$ follows also directly from a scaling
collapse of the radius distribution of dissipative first waves. 

For the distribution of duration times $t$ of avalanches exactly the
same analysis can be performed as for the number of topplings $s$. One
finds again a violation of FSS, the typical duration time of an
avalanche of radius $r$ being $\sim r^z$ with $z=4/3$, the upper
cutoff of the duration time being larger than $r^{3/2}$, and the mean
duration time of avalanches, averaged over all avalanche sizes, being
$\bar t \sim L$.

The following table summarizes the results for the exponents:

\begin{center}
\begin{tabular}{r|c|c|c|c|c}
& $\tau_r$ & $\tau_t$ & $\tau_s$ & $z$ & $D$ \\
\hline
$1^{st}$ wave & 7/4&8/5&11/8&5/4&2\\
avalanche & 3/2&11/8&11/9&4/3&9/4
\end{tabular}
\end{center}

In summary, we have shown that the scaling behavior of the abelian sandpile model in two dimensions is tied to the condition that the fraction $1/\sqrt{r}$ of all avalanches started at
a distance $r$ from the boundary reach the boundary, rather than to a
self-similarity of large and small avalanches. As a consequence, the
scaling properties are seen most clearly from the dissipative
avalanches, and a violation of FSS can occur. 

The condition that a fraction proportional to $1/\sqrt{L}$ of all
avalanches reach the boundaries, together with the two conditions
$\bar s \sim L^2$ and $\bar t \sim L$ seem so simple that they should
be derivable from simple argument. In fact, $\bar s \sim L^2$ follows
from the diffusive motion of sandgrains \cite{dha90}, and appears to
be also valid in higher dimensions \cite{gra90}. The first condition
is equivalent to the statement that the probability of triggering a
system spanning avalanche, $p$, is proportional to the density of
surface sites, $\rho$, that topple during a system-spanning
avalanche. The reason is that the product $p \rho L^{d-1}$ is the mean
number of grains dissipated during an avalanche, which must be
identical to 1, resulting in $p \sim \rho \sim 1/L^{(d-1)/2}$ ($d$
denotes the dimension of the system). A condition $p\sim \rho$ holds
for example for percolation clusters in systems above the percolation
threshold. It also holds at the percolation threshold, if there are
only a finite number of spanning clusters in the system. In three
dimensions, this condition would lead to $\tau_r=2$, which agrees with
the simulation result given in \cite{lub97}. If avalanches are
compact, as suggested by computer simulations, the exponent $\tau_a$
characterizing the area distribution in three dimensions is then
$\tau_a=4/3$, giving a mean avalanche area $\sim L^2$. Together with
the condition $\bar s = L^2$, it follows that the mean number of
topplings is of the same order as the mean area, leading to
$\tau_s=4/3$ and to the conclusion that the number of waves of
topplings remains finite in the thermodynamic limit, and that FSS is
not violated in three dimensions. Finally, a compact avalanche of
$\bar s = L^2$ sites has a radius $L^{2/d}$. In two dimensions, this
corresponds to the mean avalanche duration $\bar t$. If we assume the
same in three dimensions, we have $\bar t \sim L^{2/3}$, and we find
$\tau_t=8/5$, again in agreement with simulation results
\cite{lub97}. In dimensions above the upper critical dimension 4
\cite{pri99}, avalanches are no more compact, and the system may
contain many spanning clusters, and the above arguments can
therefore not be applied.

It thus seems that the scaling behavior of the abelian sandpile model
follows from a few simple principles, a connection which still has to
be explored in greater depth.

 \acknowledgements This work was supported by EPSRC Grant
No.~GR/K79307.

\end{multicols}

\end{document}